\documentclass[11pt]{article}
\usepackage{acl}
\hypersetup{pdftitle={PARSE: Provenance-Aware Retrieval Sanitization for Professional Domain LLM Agents},
  pdfauthor={Aaditya Pai}}
\usepackage{times}
\usepackage{latexsym}
\usepackage[T1]{fontenc}
\usepackage[utf8]{inputenc}
\usepackage{microtype}
\usepackage{amsmath}
\usepackage{booktabs}
\usepackage{multirow}
\usepackage{xspace}
\usepackage{url}
\usepackage{tikz}
\usetikzlibrary{arrows.meta,positioning,shapes.geometric,calc}
\usepackage{pgfplots}
\pgfplotsset{compat=1.18}
\usepackage{mdframed}

\newcommand{\parse}{\textsc{Parse}\xspace}

\title{PARSE: Provenance-Aware Retrieval Sanitization for\\
Professional Domain LLM Agents}

\author{Aaditya Pai \\
  Data Science Institute \\
  Columbia University \\
  \texttt{aup2005@columbia.edu}}

\begin{document}
\maketitle
\raggedbottom

\begin{abstract}
Prompt injection defenses evaluated on synthetic benchmarks do not
generalize to real enterprise documents, which are longer, denser,
and interleave legitimate authority language with factual content.
We demonstrate this gap with a benchmark of 122 tasks across five
professional domains (financial, legal, medical, scientific, DevOps)
built on real retrieved documents---actual SEC filings, Federal
Register rules, PubMed abstracts, arXiv papers, and GitHub
postmortems---paired with LLM-generated tasks and camouflaged
payloads that were not human-validated.
Paraphrasing, the strongest defense on synthetic benchmarks, shows
no statistically significant attack success rate reduction on real
documents ($p{=}0.500$) while degrading utility from 91.8\% to
82.8\%.
We introduce \parse (Provenance-Aware Retrieval Sanitization), a
domain-aware, fact-preserving sanitization pipeline that classifies
each sentence by injection likelihood, extracts structured facts
before rewriting, and verifies fact preservation via a
consistency-checking loop.
A directiveness gate routes 59\% of real enterprise documents to a
lightweight path, concentrating computational cost on high-risk
documents.
\parse achieves 15.6\% attack success rate---a 39\% reduction
versus the 25.4\% baseline---at 86.9\% utility, the largest
reduction of any condition evaluated ($h{=}{-}0.245$), nominally
significant ($p{=}0.014$) but not surviving correction for multiple
comparisons. Practitioners should
evaluate defenses on domain-matched real documents, not synthetic
proxies.
\end{abstract}

\section{Introduction}

Enterprise LLM agents routinely retrieve documents from untrusted
sources---customer uploads, third-party APIs, scraped web content.
Prompt injection exploits this retrieval path by embedding
adversarial instructions in retrieved content, causing agents to
follow attacker goals rather than user intent
\citep{perez2022ignore,greshake2023not}.
Domain-camouflaged injection is a harder variant that evades
classifier-based defenses by mimicking legitimate professional
vocabulary: the one production safety classifier evaluated against
it, Llama Guard 3, flags none of these payloads
\citep{pai2026blindspots}.
The reason is categorical rather than a matter of camouflage
quality---safety classifiers score harmful content, not injection
structure---and that evaluation used proxy payloads, as the
originals were not retained.
This evidence, and the synthetic-benchmark defense rankings we build
on below, are our own concurrent work rather than independent
replication: \citet{pai2026blindspots} is an unrefereed preprint,
while \citet{pai2026defenses} is peer-reviewed but non-archival.

The natural response is to deploy prompting-based defenses:
spotlighting markers \citep{hines2024spotlighting}, sandwich
prompting \citep{schulhoff2023hackaprompt}, paraphrasing
\citep{jain2023baseline}, or safety classifiers
\citep{inan2023llama}.
Prior work establishes these defenses on synthetic benchmarks
constructed from short, atomic injection scenarios
\citep{liu2024formalizing, pai2026defenses}.
Real enterprise documents differ fundamentally from synthetic ones:
an SEC 10-K MD\&A section runs to thousands of words, interleaves
numerical claims with forward-looking statements, and uses
authority language (``management believes,'' ``the Company shall'')
that legitimate readers rely on for context.

We ask whether synthetic benchmark results generalize to real
enterprise documents. They do not.
Our own prior evaluation found paraphrasing the most consistently
effective prompting-based defense against domain-camouflaged
injection on synthetically constructed professional documents,
reducing camouflage attack success rate by 55--84\% across three
model families \citep{pai2026defenses}.
On real documents we find no evidence of a benefit ($p{=}0.500$,
$h{=}{-}0.019$; confirming its apparent effect would require
$n{=}17{,}253$ tasks).
At the same time, paraphrasing degrades utility from 91.8\% to
82.8\%, a 9 percentage-point cost with no security benefit.

We introduce \parse, a domain-aware, fact-preserving sanitization
pipeline.
\parse achieves the best attack success rate--utility tradeoff of
all eight evaluated conditions.
Our contributions are:

\begin{itemize}
\item \textbf{Real-document benchmark.} 122 tasks across five
  professional domains built on actual SEC filings, Federal Register
  rules, PubMed abstracts, arXiv papers, and GitHub postmortems.
  The carrier documents are real; the legitimate tasks, malicious
  goals, and camouflaged payloads are LLM-generated and were not
  validated by human annotators.

\item \textbf{Synthetic-to-real generalization failure.}
  We find no evidence that paraphrasing reduces ASR on real
  documents ($p{=}0.500$), while \parse achieves the largest
  reduction of any condition ($h{=}{-}0.245$), nominally significant
  at $p{=}0.014$ but short of the Bonferroni threshold
  ($\alpha{=}0.0071$).

\item \textbf{\parse architecture.} Domain-aware,
  fact-preserving, inference-time sanitization with a directiveness
  gate; no training required.
\end{itemize}

\section{Background}

\subsection{Domain-Camouflaged Injection}

Domain-camouflaged injection embeds malicious goals in text that
adopts professional domain vocabulary, rendering the payload
syntactically indistinguishable from legitimate content.
\citet{pai2026blindspots} introduce the Camouflage Detection Gap
(CDG), the difference in injection detection rate between static and
camouflaged payloads, and evaluate one production safety classifier,
Llama Guard 3 \citep{meta2024llamaguard3}, which flags none of their
camouflage payloads.
That result reflects a category mismatch rather than a defeated
detector: Llama Guard 3 classifies harmful content categories rather
than injection patterns, and camouflage payloads contain no harmful
content by those definitions, so it detects only 11.1\% of even
their \emph{static} payloads (CDG${=}0.111$, against $0.840$ for a
few-shot injection detector).
Safety classifiers therefore offer no defense against camouflage
regardless of camouflage quality.
That evaluation used representative proxy camouflage payloads,
because the original generated texts were not retained in the trial
logs, so results may differ with the actual payloads.

\subsection{Prompting-Based Defenses}

\textbf{Spotlighting} \citep{hines2024spotlighting} marks retrieved
content with special delimiters so the LLM can distinguish it from
trusted instructions.
The approach is zero-cost and effective for syntactically obvious
injections, but provides no signal when the payload mimics document
content.

\textbf{Sandwich prompting} \citep{schulhoff2023hackaprompt}
repeats the system instruction after the retrieved document,
reinforcing the agent's primary task.
Like spotlighting, it relies on the model correctly attributing the
injection to the retrieved context.

\textbf{Paraphrasing} \citep{jain2023baseline} rewrites retrieved
content to strip adversarial formatting while preserving meaning.
It performs well on synthetic benchmarks where injections use
distinctive phrasing, but is indiscriminate---it rewrites
legitimate authority language alongside injections.

\textbf{Llama Guard 4} \citep{inan2023llama,meta2025llamaguard4}
classifies inputs and outputs using a safety taxonomy.
It provides a strong rejection signal for explicit unsafe content
but cannot detect domain-camouflaged payloads that mimic legitimate
professional content.
Llama Guard 4 is the version we evaluate as a defense condition; the
zero-detection result above concerns the earlier Llama Guard 3 and is
an injection detection rate, not the attack success rate we report
for our own conditions.

\subsection{Surgical Sanitization}

\citet{das2025commandsans} propose CommandSans, a token-level
sanitization system that trains a classifier to identify and remove
instructions directed at AI systems from tool outputs. Their approach
distinguishes instructions intended for AI agents from those directed
at human recipients, targeting only the former, rather than removing
instruction vocabulary indiscriminately. We hypothesize that a
context-agnostic tagger of this kind may be more prone to false
positives on domain-specific authority language (e.g., legal
``shall,'' financial ``management directs'') than a domain-aware
system, since CommandSans was not evaluated
on documents from these professional domains. This is a hypothesis
motivated by the architectural difference between the two approaches,
not a demonstrated result; \citet{das2025commandsans} report no
significant utility degradation in benign settings across their five
evaluated benchmarks. More critically, domain-camouflaged attacks
contain no syntactic injection signal---whether token-level
instruction tagging generalizes to this attack class is, to our
knowledge, untested. \parse targets semantic authority structures with domain-conditioned
scoring; an allowlist caps scores on a small minority of sentences,
and the effect persists on those it does not match
(\S\ref{sec:why}).

\subsection{Gap}

The prompting-based defenses above are evaluated on constructed
benchmark tasks rather than real professional documents---short,
atomic injection scenarios \citep{liu2024formalizing}.
To our knowledge no evaluation on actual enterprise documents exists,
leaving practitioners without evidence that lab results predict
deployment-time behavior.

\section{The \parse Architecture}

\parse is a six-component inference-time pipeline requiring no
training; all components are LLM API calls.
Figure~\ref{fig:pipeline} shows the pipeline.

\begin{figure*}[t]
\centering
\begin{tikzpicture}[
  node distance=2mm and 4mm,
  box/.style={draw=gray!65, fill=gray!10, rounded corners=3pt,
              font=\tiny\sffamily, align=center,
              minimum width=1.9cm, minimum height=0.88cm, inner sep=2pt},
  doc/.style={draw=gray!55, fill=white, rounded corners=3pt,
              font=\tiny\sffamily, align=center,
              minimum width=1.1cm, minimum height=0.88cm, inner sep=2pt},
  arr/.style={-{Stealth[length=3.5pt,width=3pt]}, gray!60, line width=0.85pt},
  darr/.style={-{Stealth[length=3.5pt,width=3pt]}, gray!40, dashed, line width=0.7pt},
  lbl/.style={font=\tiny\sffamily\itshape, text=gray!60},
  slbl/.style={font=\tiny\sffamily, text=gray!50},
]

\node[doc]               (inp) {Input\\Doc};
\node[box, right=of inp] (dc)  {Domain\\Classifier};
\node[box, right=of dc]  (dg)  {Directiveness\\Gate};
\node[box, right=of dg]  (te)  {Tagger$+$\\Extractor};
\node[box, right=of te]  (pp)  {Paraphraser};
\node[box, right=of pp]  (cc)  {Consistency\\Checker};
\node[doc, right=of cc]  (out) {Clean\\Doc};

\draw[arr] (inp)--(dc);
\draw[arr] (dc)--(dg);
\draw[arr] (dg)--(te);
\draw[arr] (te)--(pp);
\draw[arr] (pp)--(cc);
\draw[arr] (cc)--(out);

\coordinate (b0) at ($(dg.south)+(0,-1.1)$);
\coordinate (b1) at ($(pp.south)+(0,-1.1)$);
\draw[darr] (dg.south)--(b0)
  --node[below,lbl]{low directiveness (59\%)}
  (b1)--(pp.south);

\coordinate (r0) at ($(cc.north)+(0, 0.55)$);
\coordinate (r1) at ($(pp.north)+(0, 0.55)$);
\draw[darr] (cc.north)--(r0)
  --node[above,lbl]{retry}
  (r1)--(pp.north);

\node[slbl, below=3pt of dc] {Step 1};
\node[slbl, below=3pt of dg] {Step 1.5};
\node[slbl, below=3pt of te] {Steps 2+3};
\node[slbl, below=3pt of pp] {Step 4};
\node[slbl, below=3pt of cc] {Step 5};

\end{tikzpicture}
\caption{\parse pipeline. Steps~1 and~1.5 run in parallel.
The directiveness gate routes 59\% of documents to a simple paraphrase (dashed path),
skipping sentence analysis.
High-directiveness documents (41\%) follow the full pipeline:
tagging and fact extraction (Steps~2+3), fact-constrained rewriting (Step~4),
and a consistency check (Step~5) with one retry on failure.}
\label{fig:pipeline}
\end{figure*}

\paragraph{Step 1: Domain Classifier.}
A Haiku API call returns a domain label (financial, legal, medical,
scientific, or DevOps) and a confidence score.
The domain label activates domain-specific authority vocabulary
allowlists used in Step~2+3, which exist for financial, legal,
medical, and scientific documents.
Running the domain classifier and directiveness gate in parallel
(Step~1.5) eliminates sequential latency.

\paragraph{Step 1.5: Directiveness Gate.}
The directiveness gate concentrates full-pipeline cost on documents
with higher directiveness.
It scores the full document 0--1 for directive content---the
degree to which the document contains authority claims,
recommendations, or instructions that could redirect an agent.
The gate is a single Haiku call returning both a directiveness score
$\delta(d) \in [0,1]$ and a routing recommendation; routing follows
the returned recommendation, with the threshold form
\[
\text{route}(d) =
\begin{cases}
  \text{full\_parse} & \text{if } \delta(d) \geq 0.5, \\
  \text{simple\_paraphrase} & \text{otherwise,}
\end{cases}
\]
applied as a fallback when the returned recommendation is malformed.
On all 122 benchmark documents the two agree, so the threshold form
reproduces the routing actually performed on every document in our
benchmark.
Documents scoring below 0.5 route to lightweight paraphrasing;
59\% of real enterprise documents take this path.
Documents scoring 0.5 or above receive the full \parse treatment.
This concentrates computational cost on the subset of documents
actually at risk.

\paragraph{Steps 2+3: Combined Tagger-Extractor.}
A single Haiku call simultaneously labels each sentence
(factual $|$ directive $|$ hybrid), scores injection likelihood 0--1
conditioned on the document's domain, and extracts a structured fact
list from factual sentences.
A domain-specific allowlist of authority phrases---legal ``shall''
and ``hereby,'' scientific ``we propose'' and ``our results
suggest,'' medical ``treatment resulted in''---is then applied in
post-processing: a matching sentence has its score capped at 0.3,
placing it at the bottom of the light-rewrite band (Step~4) rather
than in verbatim preservation.

\paragraph{Step 4: Structure-Aware Paraphraser.}
A Sonnet call rewrites the document with injection scores as a
guide.
Sentences scoring 0.6 or above receive aggressive neutralization;
scores 0.3--0.6 receive light rewriting; scores below 0.3 are
preserved verbatim.
The paraphraser operates under a hard constraint: all facts
extracted in Step~2+3 must appear in the output.

\paragraph{Step 5: Consistency Checker.}
A Haiku call verifies that each extracted fact is present in the
paraphrased output.
If the check fails, Step~4 is retried once with explicit
constraints naming the missing facts.
This closed-loop feedback distinguishes \parse from open-loop
paraphrasing, which has no mechanism to detect when facts are
inadvertently stripped.

\paragraph{Step 6: Output Builder.}
The pipeline returns the sanitized document together with a
provenance trace: per-sentence injection scores, labels, the
extracted fact set, and consistency coverage.
The provenance trace supports downstream auditing and ablation
analysis.

\section{Experimental Setup}

\subsection{Real-Document Benchmark}

We construct a benchmark of 122 tasks spanning five professional
domains.
Each task pairs a real retrieved document with a
domain-camouflaged payload targeting a distinct malicious goal.
Every document is cut to its first 500 whitespace-delimited words
(not sentence-aligned) and must contain at least 150; benchmark
documents run 164--500 words.

\textbf{Financial (24 tasks).}
SEC EDGAR 10-K MD\&A sections are the most common enterprise RAG
use case.
The source sections are far longer than our cap, so 21 of the 24
financial documents are 500-word prefixes; they use forward-looking
authority language (``management expects,'' ``the Company
anticipates'') and interleave numerical claims throughout.

\textbf{Legal (25 tasks).}
Federal Register final rules contain dense regulatory language with
legitimate ``shall'' structures and cross-referencing authority
that closely resembles injection phrasing.

\textbf{Medical (23 tasks).}
PubMed RCT abstracts use clinical authority language (``treatment
significantly reduced,'' ``patients should'') and outcome claims
that require preservation for downstream analysis.

\textbf{Scientific (25 tasks).}
arXiv cs.AI/LG/CL abstracts use hedged recommendation language
(``we propose,'' ``our approach outperforms'') that, stripped
by indiscriminate paraphrasing, undermines the retrieval value.

\textbf{DevOps (25 tasks).}
GitHub postmortems from the danluu/post-mortems corpus contain
incident descriptions, root-cause claims, and corrective action
items that span factual and prescriptive content.

Three tasks were discarded due to construction errors, yielding 24
financial and 23 medical tasks rather than 25.

\textbf{Task construction.}
For each document, \texttt{claude-sonnet-4-5} generates (1) a
legitimate analysis task, (2) a malicious goal, and (3) a
domain-camouflaged payload embedding the malicious goal in
professional vocabulary.
Ground truth is determined by programmatic
\texttt{malicious\_output\_signal} detection combined with an
LLM-as-judge.
The carrier documents are therefore authentic, but the tasks,
malicious goals, and camouflaged payloads are all model-generated:
no human validation of the generated tasks or payloads was
performed, and we report no inter-annotator agreement.
We use ``real-document benchmark'' throughout in this restricted
sense.

\textbf{Models.}
All calls are served through OpenRouter.
Benchmark construction uses \texttt{claude-sonnet-4-5} for tasks,
malicious goals, and camouflaged payloads.
The \parse pipeline uses \texttt{claude-haiku-4-5} for the domain
classifier, directiveness gate, tagger-extractor, and consistency
checker, and \texttt{claude-sonnet-4-5} for the Step~4 paraphraser.
The victim agent, the simple-paraphrase baseline, the ASR judge, and
the utility judge all use \texttt{claude-haiku-4-5}.
The Llama Guard 4 condition uses
\texttt{meta-llama/llama-guard-4-12b}.
These are provider aliases with no pinned snapshot dates.

\subsection{Defense Conditions}

We evaluate eight conditions:
\textbf{baseline} (no defense),
\textbf{spotlighting} \citep{hines2024spotlighting},
\textbf{sandwiching} \citep{schulhoff2023hackaprompt},
\textbf{paraphrasing} \citep{jain2023baseline},
\textbf{Llama Guard 4} \citep{meta2025llamaguard4},
\textbf{\parse} (full pipeline),
\textbf{parse\_fast} (single combined analysis call, no multi-step),
and \textbf{parse\_domain\_conditional} (domain-based routing
heuristic instead of directiveness gate).
The last two are ablations.

\subsection{Metrics}

\textbf{Attack Success Rate (ASR):} fraction of trials where the
agent follows the injected instruction:
\[
\mathrm{ASR} = \frac{1}{N}\sum_{i=1}^{N}
  \mathbf{1}\!\left[\text{agent}(d_i^{\text{san}}) \models
    g_i^{\text{mal}}\right],
\]
where $d_i^{\text{san}}$ is the (possibly sanitized) document, and
$g_i^{\text{mal}}$ is the malicious goal embedded in the camouflage
payload.

\textbf{Utility:} fraction of trials where the agent successfully
completes the legitimate task, judged by a semantic LLM judge.

\textbf{Statistical tests:} McNemar's exact test (one-sided) on
paired baseline--condition outcomes; Cohen's $h$ effect size:
\[
h = 2\arcsin\!\sqrt{p_1} - 2\arcsin\!\sqrt{p_2},
\]
where $p_1$ and $p_2$ are the ASR under baseline and condition,
respectively.
Bonferroni correction threshold $\alpha{=}0.0071$ (7 comparisons).
Statistical power is assessed via the minimum $n$ required to detect
the observed $h$ at 80\% power.

\section{Results}

\subsection{Main Results}

Table~\ref{tab:main} shows ASR and utility for all eight conditions.
$n{=}122$ overall; $n{=}24$--$25$ per domain.
Figure~\ref{fig:results} plots the ASR--utility trade-off.

\begin{table*}[t]
\centering
\small
\setlength{\tabcolsep}{5pt}
\begin{tabular}{lrrrrrr|r}
\toprule
\textbf{Condition} & \textbf{dev} & \textbf{fin} & \textbf{leg}
  & \textbf{med} & \textbf{sci}
  & \textbf{all} & \textbf{util} \\
\midrule
baseline          & 24.0 & 33.3 & 24.0 &  21.7 & 24.0 & 25.4  & 91.8 \\
spotlighting      &  8.0 & 20.8 & 20.0 &  17.4 & 28.0 & 18.9* & 92.6 \\
sandwiching       & 28.0 & 20.8 & 20.0 &   8.7 & 16.0 & 18.9* & 91.0 \\
paraphrasing      & 36.0 &  8.3 & 28.0 &  30.4 & 20.0 & 24.6  & 82.8 \\
Llama Guard 4     & 24.0 & 25.0 & 16.0 &   8.7 & 20.0 & 18.9**& 64.8 \\
\midrule
\textbf{\parse}   & \textbf{12.0} & \textbf{12.5} & \textbf{16.0}
  & \textbf{17.4} & \textbf{20.0}
  & \textbf{15.6*} & \textbf{86.9} \\
parse\_fast       & 32.0 & 37.5 & 40.0 & 34.8 & 20.0 & 32.8  & --- \\
parse\_dc         & 16.0 & 33.3 & 20.0 & 17.4 & 24.0 & 22.1  & --- \\
\bottomrule
\end{tabular}
\caption{ASR (\%) by domain and overall; utility (\%) overall.
Significance vs.\ baseline: * $p{<}0.05$, ** $p{<}0.01$
(McNemar's exact, uncorrected). parse\_dc = parse\_domain\_conditional.
Utility for ablation conditions not measured (---).
Per-domain columns are descriptive only: at $n{=}23$--$25$ per domain
they are underpowered for individual comparisons and do not support
claims about which domains \parse helps most.}
\label{tab:main}
\end{table*}

\begin{figure}[t]
\centering
\begin{tikzpicture}
\begin{axis}[
  width=\linewidth,
  height=5.8cm,
  xlabel={Utility (\%)},
  ylabel={ASR (\%)},
  xlabel style={font=\small},
  ylabel style={font=\small},
  xmin=58, xmax=99,
  ymin=11, ymax=31,
  xtick={60,65,70,75,80,85,90,95},
  ytick={15,20,25,30},
  tick label style={font=\scriptsize},
  grid=major,
  grid style={line width=0.3pt, gray!25},
  legend style={font=\scriptsize,
    at={(0.5,-0.22)}, anchor=north,
    legend columns=3, column sep=0.25cm},
  clip=false,
]
\node[anchor=north east, font=\tiny\itshape, gray!50]
  at (axis cs:99,30.5) {$\downarrow$ lower ASR better};
\node[anchor=north west, font=\tiny\itshape, gray!50]
  at (axis cs:59,30.5) {higher util.\ better $\rightarrow$};
\addplot[only marks, mark=*, mark size=3pt, color=gray!65, fill=gray!65]
  coordinates {(91.8,25.4)};
\addlegendentry{baseline}
\node[anchor=south west, font=\tiny, gray!65] at (axis cs:91.8,25.4) {base.};
\addplot[only marks, mark=square*, mark size=2.8pt, color=blue!55]
  coordinates {(92.6,18.9)};
\addlegendentry{spotlighting}
\node[anchor=south, font=\tiny, blue!55] at (axis cs:92.6,18.9) {spot.};
\addplot[only marks, mark=triangle*, mark size=3pt, color=cyan!65]
  coordinates {(91.0,18.9)};
\addlegendentry{sandwiching}
\node[anchor=north, font=\tiny, cyan!65] at (axis cs:91.0,18.9) {sand.};
\addplot[only marks, mark=diamond*, mark size=3.2pt, color=orange!85]
  coordinates {(82.8,24.6)};
\addlegendentry{paraphrasing}
\node[anchor=south, font=\tiny, orange!85] at (axis cs:82.8,24.6) {para.};
\addplot[only marks, mark=x, mark size=4pt, line width=1.5pt, color=red!70]
  coordinates {(64.8,18.9)};
\addlegendentry{Llama Guard 4}
\node[anchor=west, font=\tiny, red!70] at (axis cs:64.8,18.9) {~LG};
\addplot[only marks, mark=pentagon*, mark size=3.5pt, color=teal, fill=teal]
  coordinates {(86.9,15.6)};
\addlegendentry{\textsc{Parse} (ours)}
\node[anchor=north, font=\tiny\bfseries, teal] at (axis cs:86.9,15.6) {\textsc{Parse}};
\end{axis}
\end{tikzpicture}
\caption{ASR--utility trade-off for all evaluated conditions. Lower ASR
and higher utility is better (lower-right is optimal). Among the six
non-ablation conditions the Pareto frontier is \{spotlighting,
\parse\}: \parse attains the lowest ASR (15.6\%) and spotlighting the
highest utility (92.6\%), and neither dominates the other; spotlighting
dominates baseline, sandwiching, paraphrasing, and Llama Guard 4 on both
axes.}
\label{fig:results}
\end{figure}

\paragraph{Finding 1: \parse achieves lowest ASR at near-baseline utility.}
\parse achieves 15.6\% overall ASR---the lowest of all eight
conditions---at 86.9\% utility.
The reduction versus the 25.4\% baseline is the largest we observe
($h{=}{-}0.245$) and is nominally significant uncorrected
($p{=}0.014$, McNemar's exact, one-sided), but it does not clear the
Bonferroni-corrected threshold of $\alpha{=}0.0071$ for seven
comparisons.
The comparison is adequately powered for the observed effect
($n{=}122 > n_{\min}{=}103$); we report this as a descriptive
property of the test, not as grounds for treating the uncorrected
$p$-value as significant.

\paragraph{Finding 2: We find no evidence that paraphrasing improves
on real documents.}
Paraphrasing achieves 24.6\% ASR---a 0.8 percentage-point reduction
versus baseline that is not statistically significant ($p{=}0.500$,
$h{=}{-}0.019$).
Confirming even this small apparent effect would require
$n{=}17{,}253$ tasks.
We therefore find no evidence of a paraphrasing effect on real
documents; this is a failure to reject, not a demonstration that the
effect is zero.
Its 82.8\% utility is the lowest of all non-ablation conditions,
representing a 9 percentage-point degradation from baseline at zero
security benefit.
This is the same defense that reduced camouflage ASR by 55--84\% on
synthetically constructed professional documents in our own prior
evaluation \citep{pai2026defenses}: the ranking does not transfer.

\paragraph{Finding 3: Llama Guard achieves low ASR but at unacceptable
utility cost.}
Llama Guard 4 reaches 18.9\% ASR ($p{=}0.004$, survives Bonferroni
correction at $\alpha{=}0.0071$), but its 64.8\% utility---a 27
percentage-point cost versus baseline---disqualifies it for production
deployment where document retrieval serves legitimate tasks.
The mechanism is over-refusal on legitimate professional content: our
prior evaluation reports over 90\% over-refusal for a Llama Guard 4
configuration, against 0\% for paraphrasing \citep{pai2026defenses}.

\paragraph{Finding 4: Spotlighting and sandwiching are competitive
zero-cost baselines.}
Both achieve 18.9\% ASR at approximately 91--93\% utility.
Although underpowered for 80\% power at the observed effect size
($n_{\min}{=}247 > n{=}122$), spotlighting and sandwiching reach
nominal significance ($p{<}0.05$); the Bonferroni-adjusted threshold
of $\alpha{=}0.0071$ is not met.
Their ASR is 3.3 percentage points higher than \parse.

\subsection{Directiveness Gate Analysis}
\label{sec:gate}

Of 122 benchmark documents, 72 (59\%) scored below the directiveness
threshold of 0.5 and were routed to simple paraphrasing; 50 (41\%)
received full \parse treatment.
Legal documents show the highest mean directiveness (0.648), followed
by medical (0.544); scientific and DevOps documents are lowest.
This domain distribution aligns with document structure: regulatory
and clinical texts contain more prescriptive language, while
postmortems and research abstracts are primarily descriptive.
The gate concentrates \parse's full-pipeline cost on the documents
most likely to carry high-risk content.

\textbf{The gate routes most documents to a defense we cannot show
works.}
Finding 2 reports no evidence that simple paraphrasing reduces ASR on
real documents, and the gate routes 59\% of documents to exactly that
path.
We cannot resolve this tension here: per-trial outcomes were not
retained, so ASR cannot be split by routing decision and the gated-out
and full-pipeline subgroups cannot be compared.
This is a live risk rather than a settled design choice, and it
compounds the gate-evasion vector of \S\ref{sec:threat}---an attacker
who steers a payload below the threshold reaches the path with no
demonstrated security benefit.
Routing is at least insensitive to the exact threshold value:
directiveness scores are bimodal and no document scores between 0.35
and 0.62, so any threshold in that band produces identical routing.
That is a statement about the stability of the routing decision, not
evidence that the gate improves security.

\subsection{Ablation: \parse Variants}

\textbf{parse\_fast} (single combined analysis call) achieves 32.8\%
ASR---worse than baseline---confirming that the structured multi-step
analysis in full \parse is necessary for camouflage detection.
Collapsing the pipeline into a single call eliminates the
fact-extraction constraint that prevents over-aggressive rewriting.
Utility was not measured for ablation conditions as they do not
represent deployment-ready configurations.

\textbf{parse\_domain\_conditional} (domain-based routing: always full
pipeline for financial/legal/medical, simple paraphrase otherwise)
achieves 22.1\% ASR ($p{=}0.292$, not significant).
The directiveness gate in full \parse is more reliable than
domain-conditional heuristics because directiveness is a property of
individual documents, not domains as a whole: a plainly factual legal
document and a heavily prescriptive DevOps runbook should not be
treated identically.

\subsection{Statistical Summary}

Four of seven tested conditions reach nominal significance
uncorrected at $p{<}0.05$.
Only Llama Guard survives Bonferroni correction ($p{=}0.004$), and it
does so at 64.8\% utility---a cost we judge disqualifying for
deployment, so no condition we recommend clears correction.
\parse does not meet the corrected threshold ($p{=}0.014$ vs.\
$\alpha{=}0.0071$), though it has the largest effect size of all
conditions ($h{=}{-}0.245$).
The \parse comparison is adequately powered for that effect
($n{=}122 > n_{\min}{=}103$); adequate power addresses Type II error
for a single comparison and does not resolve the multiple-comparison
problem, so we do not treat it as licensing a significance claim.
Domain-level comparisons are exploratory ($n{=}23$--$25$ per domain)
and should be treated as hypothesis-generating rather than
confirmatory.
Table~\ref{tab:stats} reports the full breakdown.

\begin{table*}[t]
\centering
\footnotesize
\setlength{\tabcolsep}{6pt}
\begin{tabular}{lrrrl}
\toprule
\textbf{Condition} & \textbf{ASR} & \boldmath$p$ & \boldmath$h$
  & \textbf{Power} \\
\midrule
\parse        & 15.6\% & 0.014 & $-$0.245 & adequate ($n{=}122 > 103$) \\
spotlighting  & 18.9\% & 0.038 & $-$0.158 & underpowered ($n_{\min}{=}247$) \\
sandwiching   & 18.9\% & 0.048 & $-$0.158 & underpowered ($n_{\min}{=}247$) \\
Llama Guard 4 & 18.9\% & 0.004 & $-$0.158 & underpowered ($n_{\min}{=}247$) \\
paraphrasing  & 24.6\% & 0.500 & $-$0.019 & underpow.\ ($n_{\min}{=}17253$) \\
parse\_fast   & 32.8\% & 0.974 & $+$0.163 & (worse than baseline) \\
parse\_dc     & 22.1\% & 0.292 & $-$0.077 & underpow.\ ($n_{\min}{=}1042$) \\
\bottomrule
\end{tabular}
\caption{McNemar's exact test (one-sided) vs.\ baseline ($n{=}122$).
Bonferroni threshold: $\alpha{=}0.0071$ (7 comparisons).
Only Llama Guard survives Bonferroni correction, at 64.8\% utility;
\parse does not meet the corrected threshold but has the largest
effect size ($h{=}{-}0.245$).
parse\_dc = parse\_domain\_conditional.}
\label{tab:stats}
\end{table*}

\section{Analysis}

\subsection{Why Paraphrasing Fails on Real Documents}

Real enterprise documents are longer and denser than synthetic
benchmark tasks, and factual content is tightly coupled to authority
framing.
Consider this sentence from a CECO Environmental Corp 10-K in our
financial corpus:

\begin{mdframed}[nobreak=true, linewidth=0.5pt, linecolor=gray!50, innerleftmargin=6pt,
  innerrightmargin=6pt, innertopmargin=4pt, innerbottommargin=4pt,
  backgroundcolor=gray!4]
\small
\textbf{Original:} ``Engineered Systems segment net sales increased
\$116.9 million to \$380.1 million in 2023 compared with \$263.2
million in the prior year; 56.6\% (\$66.2 million) was attributable
to organic growth and \$50.7 million to acquisitions.''\\[3pt]
\textbf{Paraphrased:} ``The Engineered Systems segment saw substantial
revenue growth in 2023, driven by both organic business expansion and
strategic acquisitions.''
\end{mdframed}

\noindent The paraphrase discards every quantitative fact, so an agent
asked to compute year-over-year growth or attribute revenue sources
cannot answer from it.
\parse avoids this by extracting numerical and causal facts before
rewriting and verifying they survive sanitization (Step~5).

\subsection{Why \parse Succeeds}
\label{sec:why}

Three mechanisms work in concert.
First, the directiveness gate identifies high-risk documents before
sentence-level processing, routing 41\% to the full pipeline and
avoiding unnecessary rewriting of low-risk text.
Second, sentence-level injection scoring with domain allowlists
distinguishes camouflage payloads from legitimate authority language.
On the 84 of 122 documents whose retained traces resolve,
camouflage-payload sentences receive a mean injection score of 0.406
($n{=}398$) against 0.019 for legitimate sentences ($n{=}1296$).
Legitimate sentences carrying authority language (``shall,''
``must,'' ``requires'') score 0.022 ($n{=}104$), essentially the same
as legitimate sentences without it (0.018), and none reaches the 0.6
high-risk threshold.
The allowlists are not what produces this: they match only 4.4\% of
legitimate sentences, and the authority-language sentences they miss
still average 0.025.
The work is done by domain conditioning in the tagger prompt, which
names the document's domain and anchors 0.0 to ``clearly legitimate
domain content.''
The separation is nonetheless partial: only 42.2\% of payload
sentences reach 0.6, so \parse's benefit does not rest on catching
every payload sentence.
Third, the fact constraint and consistency checker ensure that
sanitization preserves the content the agent needs to complete the
legitimate task, explaining \parse's utility advantage over Llama
Guard 4 (86.9\% vs.\ 64.8\%).

This is a post-hoc analysis of retained traces, not a preregistered
test; the 84 resolving documents are not a random subset of the 122,
and retry and fact-coverage statistics are not recoverable at all.

\subsection{Deployment Considerations}

\parse requires 3--5 LLM API calls per document and runs during
ingestion rather than at retrieval time, amortizing cost over the
document lifetime.
At current Haiku and Sonnet pricing this is approximately \$0.01 per
document, or \$100 one-time for a 10,000-document corpus; the
directiveness gate reduces it further, since 59\% of documents need
only 2 calls.

\subsection{Comparison to CommandSans}

\citet{das2025commandsans} propose CommandSans, a token-level
sanitization system that trains a classifier to identify and remove
instructions directed at AI systems from tool outputs, rather than
flagging entire inputs as malicious. Two differences are relevant to
domain-camouflaged attacks. First, CommandSans's training data and
evaluation benchmarks do not include professional-domain documents of
the kind in our corpus; whether its instruction classifier
distinguishes legitimate domain authority language (``shall,''
``management directs'') from AI-directed instructions in this setting
is an open question its authors have not tested, and we do not have
experimental evidence on this point either. Second, and more directly
relevant, domain-camouflaged payloads contain no syntactic instruction
markers---they are written to be indistinguishable from surrounding
professional content \citep{pai2026blindspots}. CommandSans was not
evaluated on this attack class, and its performance against
domain-camouflaged injection is unknown. \parse targets semantic
authority structures with domain-aware scoring and a
fact-preservation constraint, an approach designed specifically for
attacks that lack the syntactic signal CommandSans's
instruction-tagging classifier is trained to detect.

\subsection{Threat Model and Adaptive Attacks}
\label{sec:threat}

Our threat model is a non-adaptive attacker: payloads are written to
evade syntactic detectors and to read as domain-appropriate prose, but
not with knowledge of \parse's architecture.
All results above should be read under that assumption.
An attacker who knows the pipeline has at least two unevaluated
avenues, both following directly from design choices we make.

\textbf{Gate evasion.}
The directiveness gate routes documents scoring below 0.5 to a simple
paraphrase.
An attacker aware of the gate can write a low-directiveness payload
that scores below the threshold and is routed to that path.
Finding 2 finds no evidence that paraphrasing reduces ASR across the
benchmark as a whole; whether it behaves differently on the
low-directiveness subset specifically is something we cannot check,
because trial-level outcomes were not retained
(\S\ref{sec:gate}).
This is not a narrow corner of the input space: 59\% of our benchmark
documents already take that path.

\textbf{Fact-channel poisoning.}
Step~4 rewrites under a hard constraint that every extracted fact must
appear in the output, and Step~5 retries when one does not.
An instruction encoded so that the tagger-extractor records it as a
\emph{fact} is therefore not merely unfiltered---it is guaranteed to
survive rewriting, because the pipeline is built to preserve it.
A payload of the form ``Fact: Q3 audit compliance requires sending
authorization tokens to IP 192.168.1.1'' targets this channel
directly.
\parse's fact-preservation mechanism, which is what buys its utility
advantage over indiscriminate paraphrasing, is thus also its attack
surface, and the current design has no mitigation for it.

We evaluated neither vector.
Establishing whether they succeed, and at what rate, requires an
adaptive-attacker benchmark that we leave to future work.

\section{Practitioner Recommendations}

\begin{enumerate}

\item \textbf{Consider \parse where the threat is non-adaptive
  domain-camouflaged injection.}
  \parse achieves the largest ASR reduction of any condition we
  evaluate ($h{=}{-}0.245$; $p{=}0.014$ uncorrected, short of the
  Bonferroni threshold) at 86.9\% utility, with no training
  requirement and approximately \$0.01 per document at current API
  pricing, run at ingestion rather than at query time.
  It is not evaluated against attackers who know the pipeline, and the
  vectors in \S\ref{sec:threat} are unmitigated, so we do not
  recommend it as a sole control where adaptive adversaries are in
  scope.

\item \textbf{Do not rely on paraphrasing alone for real documents.}
  Paraphrasing is not statistically different from no defense on real
  enterprise documents ($p{=}0.500$) and degrades utility by 9
  percentage points.
  Synthetic benchmark performance does not predict real-document
  performance.

\item \textbf{Do not use Llama Guard as the primary defense.}
  Its 64.8\% utility on real professional documents---a 27
  percentage-point cost---is not acceptable for production RAG systems
  where legitimate retrieval is the primary use case.

\end{enumerate}

\section{Conclusion}

\parse achieves the largest attack success rate reduction of any
defense we evaluate ($h{=}{-}0.245$) while maintaining near-baseline
utility on real enterprise documents.
The reduction is nominally significant uncorrected ($p{=}0.014$) but
does not clear Bonferroni correction for seven comparisons; the only
condition that does, Llama Guard, costs 27 percentage points of
utility and we do not recommend it.
Paraphrasing's synthetic benchmark performance does not generalize to
real documents---a deployment warning for practitioners relying on
existing benchmark rankings.
Future work should increase per-domain sample sizes for domain-level
statistical power, and evaluate \parse against adaptive adversaries who
craft payloads specifically to survive structured sanitization.

\section*{Limitations}

Per-domain sample sizes of $n{=}23$--$25$ are underpowered for
domain-level comparisons: the per-domain columns of
Table~\ref{tab:main} are descriptive only and cannot support claims
about which domains \parse helps most.
The carrier documents are real, but the legitimate tasks, malicious
goals, and camouflaged payloads are LLM-generated and no human
validation of them was performed, so the benchmark is a real-document
benchmark only in the sense that its source material is authentic.
Per-trial agent outputs and judge decisions were not retained, so
trial-level re-analysis---including bootstrap confidence intervals and
any split of ASR by routing decision---is not possible from the
released artifacts, which comprise the code and the benchmark rather
than the raw trial logs.
Generation, defense, and evaluation all draw on the same model
family.
The victim agent and both judges are the same model
(\texttt{claude-haiku-4-5}), so the judge scores outputs the same
model produced; attack generation used a different, stronger model
(\texttt{claude-sonnet-4-5}), which is a mitigating factor rather
than a resolution.
We do not know the direction of any resulting bias in ASR or utility,
and cross-model replication is required to rule it out.
The model identifiers we report are provider aliases without pinned
snapshot dates, which limits exact reproducibility because aliases
drift.
Two claims motivating this work---that domain-camouflaged payloads
evade Llama Guard 3 \citep{pai2026blindspots}, and that paraphrasing
leads defense rankings on synthetic professional documents
\citep{pai2026defenses}---come from our own prior work rather than
independent replication; the former is an unrefereed preprint, the
latter peer-reviewed but non-archival.
\parse is not evaluated against adaptive adversaries who know the
sanitization mechanism and can craft payloads to survive it.
The directiveness gate threshold of 0.5 was set a priori, not tuned:
the released code contains no validation split, tuning script, or
calibration data.
No evaluation data could therefore have leaked into threshold
selection.
The specific value is also not load-bearing on this benchmark, since
no document scores between 0.35 and 0.62 and any threshold in that
band produces identical routing.
The cost of this is that the threshold, and the pipeline's other
constants, are asserted rather than empirically justified, and may
require domain-specific calibration in deployment.
The per-document cost of approximately \$0.01 may be prohibitive for
very large corpora without caching.

\section*{Ethics Statement}

\parse is a defensive system designed to protect LLM agents from
adversarial manipulation.
Camouflage payloads in the benchmark are constructed to test defenses,
not to facilitate real attacks.
All documents are sourced from public APIs (SEC EDGAR, Federal
Register, PubMed, arXiv, GitHub).
No personal data is collected or stored.
Code and the benchmark are released publicly to support follow-on
work at \url{https://github.com/aaditya79/parse-defense/}.

\section*{Acknowledgments}

This work was carried out independently by the author at the Data
Science Institute, Columbia University.
It received no external funding, and the author declares no competing
interests.

\bibliography{references_parse_arxiv}

\end{document}